\documentstyle[12pt,twoside]{article}

\def\a{\alpha}
\def\b{\beta}

\def\d{\delta}
\def\e{\epsilon}                
\def\f{\phi}                    
\def\g{\gamma}
\def\h{\eta}

\def\j{\psi}

\def\l{\lambda}
\def\m{\mu}
\def\n{\nu}

\def\p{\pi}                     
\def\r{\rho}                    

\def\F{\Phi}
\def\G{\Gamma}

\def\L{\Lambda}

\def\S{\Sigma}

\def\cf{{\cal F}}

\def\ch{{\cal H}}

\def\cn{{\cal N}}

\def\cs{{\cal S}}

\catcode`@=11

\def\un#1{\relax\ifmmode\@@underline#1\else $\@@underline{\hbox{#1}}$\relax\fi}

{\catcode`\'=\active \gdef'{{}~\bgroup\prim@s}}

\def\magstep#1{\ifcase#1 \@m\or 1200\or 1440\or 1728\or 2074\or 2488\or
        2986\fi\relax}

\font\twfvmi=cmmi10\@magscale5
    \skewchar\twfvmi='177
\font\twfvsy=cmsy10\@magscale5
    \skewchar\twfvsy='60
\font\twfvly=lasy10\@magscale5
\font\thtyrm=cmr10\@magscale6

\def\vpt{\textfont\z@\fivrm
  \scriptfont\z@\fivrm \scriptscriptfont\z@\fivrm
\textfont\@ne\fivmi \scriptfont\@ne\fivmi \scriptscriptfont\@ne\fivmi
\textfont\tw@\fivsy \scriptfont\tw@\fivsy \scriptscriptfont\tw@\fivsy
\textfont\thr@@\tenex \scriptfont\thr@@\tenex \scriptscriptfont\thr@@\tenex
\def\prm{\fam\z@\fivrm}%
\def\unboldmath{\everymath{}\everydisplay{}\@nomath
  \unboldmath\fam\@ne\@boldfalse}\@boldfalse
\def\boldmath{\@subfont\boldmath\unboldmath}%
\def\pit{\@getfont\pit\itfam\@vpt{cmti5}}%
\def\psl{\@subfont\sl\it}%
\def\pbf{\@getfont\pbf\bffam\@vpt{cmbx5}}%
\def\ptt{\@subfont\tt\rm}%
\def\psf{\@subfont\sf\rm}%
\def\psc{\@subfont\sc\rm}%
\def\ly{\fam\lyfam\fivly}\textfont\lyfam\fivly
    \scriptfont\lyfam\fivly \scriptscriptfont\lyfam\fivly
\@setstrut\rm}

\def\@vpt{}

\def\vipt{\textfont\z@\sixrm
  \scriptfont\z@\sixrm \scriptscriptfont\z@\sixrm
\textfont\@ne\sixmi \scriptfont\@ne\sixmi \scriptscriptfont\@ne\sixmi
\textfont\tw@\sixsy \scriptfont\tw@\sixsy \scriptscriptfont\tw@\sixsy
\textfont\thr@@\tenex \scriptfont\thr@@\tenex \scriptscriptfont\thr@@\tenex
\def\prm{\fam\z@\sixrm}%
\def\unboldmath{\everymath{}\everydisplay{}\@nomath
  \unboldmath\@boldfalse}\@boldfalse
\def\boldmath{\@subfont\boldmath\unboldmath}%
\def\pit{\@subfont\it\rm}%
\def\psl{\@subfont\sl\rm}%
\def\pbf{\@getfont\pbf\bffam\@vipt{cmbx6}}%
\def\ptt{\@subfont\tt\rm}%
\def\psf{\@subfont\sf\rm}%
\def\psc{\@subfont\sc\rm}%
\def\ly{\fam\lyfam\sixly}\textfont\lyfam\sixly
    \scriptfont\lyfam\sixly \scriptscriptfont\lyfam\sixly
\@setstrut\rm}

\def\@vipt{}

\def\xxxpt{\textfont\z@\thtyrm
  \scriptfont\z@\twfvrm \scriptscriptfont\z@\twtyrm
\textfont\@ne\twfvmi \scriptfont\@ne\twfvmi \scriptscriptfont\@ne\twtymi
\textfont\tw@\twfvsy \scriptfont\tw@\twfvsy \scriptscriptfont\tw@\twtysy
\textfont\thr@@\tenex \scriptfont\thr@@\tenex \scriptscriptfont\thr@@\tenex
\def\unboldmath{\everymath{}\everydisplay{}\@nomath\unboldmath
        \textfont\@ne\twfvmi \textfont\tw@\twfvsy \textfont\lyfam\twfvly
        \@boldfalse}\@boldfalse
\def\boldmath{\@subfont\boldmath\unboldmath}%
\def\prm{\fam\z@\thtyrm}%
\def\pit{\@subfont\it\rm}%
\def\psl{\@subfont\sl\rm}%
\def\pbf{\@getfont\pbf\bffam\@xxxpt{cmbx10\@magscale6}}%
\def\ptt{\@subfont\tt\rm}%
\def\psf{\@subfont\sf\rm}%
\def\psc{\@subfont\sc\rm}%
\def\ly{\fam\lyfam\twfvly}\textfont\lyfam\twfvly
   \scriptfont\lyfam\twfvly \scriptscriptfont\lyfam\twtyly
\@setstrut \rm}

\def\@xxxpt{}

\def\Huge{\@setsize\Huge{36pt}\xxxpt\@xxxpt}

\font\thtymi=cmmi10\@magscale6
    \skewchar\thtymi='177
\font\thtysy=cmsy10\@magscale6
    \skewchar\thtysy='60
\font\thtyly=lasy10\@magscale6
\font\thsirm=cmr12\@magscale6

\def\xxxvipt{\textfont\z@\thsirm
  \scriptfont\z@\thtyrm \scriptscriptfont\z@\twfvrm
\textfont\@ne\thtymi \scriptfont\@ne\thtymi \scriptscriptfont\@ne\twfvmi
\textfont\tw@\thtysy \scriptfont\tw@\thtysy \scriptscriptfont\tw@\twfvsy
\textfont\thr@@\tenex \scriptfont\thr@@\tenex \scriptscriptfont\thr@@\tenex
\def\unboldmath{\everymath{}\everydisplay{}\@nomath\unboldmath
        \textfont\@ne\thtymi \textfont\tw@\thtysy \textfont\lyfam\thtyly
        \@boldfalse}\@boldfalse
\def\boldmath{\@subfont\boldmath\unboldmath}%
\def\prm{\fam\z@\thsirm}%
\def\pit{\@subfont\it\rm}%
\def\psl{\@subfont\sl\rm}%
\def\pbf{\@getfont\pbf\bffam\@xxxpt{cmss12\@magscale6}}%
\def\ptt{\@subfont\tt\rm}%
\def\psf{\@subfont\sf\rm}%
\def\psc{\@subfont\sc\rm}%
\def\ly{\fam\lyfam\thtyly}\textfont\lyfam\thtyly
   \scriptfont\lyfam\thtyly \scriptscriptfont\lyfam\twfvly
\@setstrut \rm}

\def\@xxxvipt{}

\def\HUGE{\@setsize\HUGE{43pt}\xxxvipt\@xxxvipt}

\catcode`@=12

\font\tenex=cmex10 scaled 1200

\def\Sc#1{\hbox{\sc #1}}        

\def\bo{{\raise.05ex\hbox{\large$\Box$}\:}}             
\def\cbo{{\,\raise-.15ex\Sc [\,}}                       
\def\pa{\partial}                                       
\def\su{\sum}                                           
\def\TH{{\raise.2ex\hbox{$\displaystyle \bigodot$}\mskip-4.7mu \llap H \;}}
\def\face{\hbox{\normalsize$\;\;\:{\raise.9ex\hbox{\oo n}\mskip-13mu \llap
        {${\buildrel{\hbox{\frtnrm ..}}\over\smile}$}}\:$}}     
\def\Face{{\raise.2ex\hbox{$\displaystyle \bigodot$}\mskip-2.2mu \llap {$\ddot
        \smile$}}}                                      
\def\Lhat{{\bf\rlap{\kern-.09em$\hat{\phantom L}$}L}}
\def\Lcheck{{\bf\rlap{\kern-.09em$\check{\phantom L}$}L}}

\def\sp#1{{}^{#1}}                              
\def\sb#1{{}_{#1}}                              
\def\sl#1{\rlap{\hbox{$\mskip 1 mu /$}}#1}      
\def\sbra#1{\left\langle #1\right|}             
\def\sket#1{\left| #1\right\rangle}             
\def\leftrightarrowfill{$\mathsurround=0pt \mathord\leftarrow \mkern-6mu
        \cleaders\hbox{$\mkern-2mu \mathord- \mkern-2mu$}\hfill
        \mkern-6mu \mathord\rightarrow$}
\def\dvec#1{\vbox{\ialign{##\crcr
        \leftrightarrowfill\crcr\noalign{\kern-1pt\nointerlineskip}
        $\hfil\displaystyle{#1}\hfil$\crcr}}}           
\def\dt#1{{\buildrel {\hbox{\LARGE .}} \over {#1}}}     
\def\ddt#1{{\buildrel {\hbox{\LARGE .\kern-2pt.}} \over {#1}}}

\def\frac#1#2{{\textstyle{#1\over\vphantom2\smash{\raise.20ex
        \hbox{$\scriptstyle{#2}$}}}}}                   
\def\ha{\frac12}                                        
\def\sfrac#1#2{{\vphantom1\smash{\lower.5ex\hbox{\small$#1$}}\over
        \vphantom1\smash{\raise.4ex\hbox{\small$#2$}}}} 
\def\bfrac#1#2{{\vphantom1\smash{\lower.5ex\hbox{$#1$}}\over
        \vphantom1\smash{\raise.3ex\hbox{$#2$}}}}       
\def\afrac#1#2{{\vphantom1\smash{\lower.5ex\hbox{$#1$}}\over#2}}    

\def\boxes#1{
        \newcount\num
        \num=1
        \newdimen\downsy
        \downsy=-1.64ex
        \mskip-7.8mu
        \bo
        \loop
        \ifnum\num<#1
        \llap{\raise\num\downsy\hbox{$\bo$}}
        \advance\num by1
        \repeat}
\def\boxup#1#2{\newcount\numup
        \numup=#1
        \advance\numup by-1
        \newdimen\upsy
        \upsy=.82ex
        \mskip7.8mu
        \raise\numup\upsy\hbox{$#2$}}

\newskip\humongous \humongous=0pt plus 1000pt minus 1000pt
\def\caja{\mathsurround=0pt}

\newif\ifdtup
\def\panorama{\global\dtuptrue \openup2\jot \caja
        \everycr{\noalign{\ifdtup \global\dtupfalse
        \vskip-\lineskiplimit \vskip\normallineskiplimit
        \else \penalty\interdisplaylinepenalty \fi}}}
\def\li#1{\panorama \tabskip=\humongous                         
        \halign to\displaywidth{\hfil$\displaystyle{##}$
        \tabskip=0pt&$\displaystyle{{}##}$\hfil
        \tabskip=\humongous&\llap{$##$}\tabskip=0pt
        \crcr#1\crcr}}

\def\NP{Nucl. Phys. B}
\def\PL{Phys. Lett. }

\def\PRD{Phys. Rev. D}
\def\ref#1{$\sp{#1]}$}

\parskip=\medskipamount                 
\def\baselinestretch{1.2}       
\thicklines                         
\def\title#1#2#3#4{
\begin{document}
        {\hbox to\hsize{#4 \hfill Imperial-TP/92-93/08 \& QMW/PH/ #3}}\par
        \begin{center}\vskip.5in minus.1in {\Large\bf #1}\\[.5in minus.2in]{#2}
        \vskip1.4in minus1.2in {\bf ABSTRACT}\\[.1in]\end{center}
        \begin{quotation}\par}
\def\author#1#2{#1\\[.1in]{\it #2}\\[.1in]}

\def\AM{Aleksandar Mikovi\'c
\footnote{E-mail address: UMAPT66@VAXA.CC.IC.AC.UK}
\\[.1in]{\it Department of Physics, Queen Mary and Westfield
College\\Mile End Road, London E1 4NS, U.K. }\\[.1in]and
\\[.1in] {\it Blackett Laboratory, Imperial
College\\ Prince Consort Road, London SW7 2BZ, U.K.}\footnote{present
address} \\[.1in]}

\def\WS{W. Siegel\\[.1in] {\it Institute for Theoretical
        Physics,\\ State University of New York, Stony Brook, NY 11794-3840}
        \\[.1in]}

\def\endtitle{\par\end{quotation}\vskip3.5in minus2.3in\newpage}


\def\endabstract{\par\end{quotation}
        \renewcommand{\baselinestretch}{1.2}\small\normalsize}


\def\xpar{\par}                                         
\def\letterhead{
        \centerline{\large\sf IMPERIAL COLLEGE}
        \centerline{\sf Blackett Laboratory}
        \vskip-.07in
        \centerline{\sf Prince Consort Road, London SW7 2BZ}
        \rightline{\scriptsize\sf Dr. Aleksandar Mikovi\'c}
        \vskip-.07in
        \rightline{\scriptsize\sf Tel: 071-589-5111/6983}
        \vskip-.07in
        \rightline{\scriptsize\sf E-mail: UMAPT66@VAXA.CC.IC.AC.UK}}
\def\sig#1{{\leftskip=3.75in\parindent=0in\goodbreak\bigskip{Sincerely yours,}
\nobreak\vskip .7in{#1}\par}}


\def\ree#1#2#3{
        \hfuzz=35pt\hsize=5.5in\textwidth=5.5in
        \begin{document}
        \ttraggedright
        \par
        \noindent Referee report on Manuscript \##1\\
        Title: #2\\
        Authors: #3}


\def\start#1{\pagestyle{myheadings}\begin{document}\thispagestyle{myheadings}
        \setcounter{page}{#1}}


\catcode`@=11

\def\ps@myheadings{\def\@oddhead{\hbox{}\footnotesize\bf\rightmark \hfil
        \thepage}\def\@oddfoot{}\def\@evenhead{\footnotesize\bf
        \thepage\hfil\leftmark\hbox{}}\def\@evenfoot{}
        \def\sectionmark##1{}\def\subsectionmark##1{}
        \topmargin=-.35in\headheight=.17in\headsep=.35in}
\def\ps@acidheadings{\def\@oddhead{\hbox{}\rightmark\hbox{}}
        \def\@oddfoot{\rm\hfil\thepage\hfil}
        \def\@evenhead{\hbox{}\leftmark\hbox{}}\let\@evenfoot\@oddfoot
        \def\sectionmark##1{}\def\subsectionmark##1{}
        \topmargin=-.35in\headheight=.17in\headsep=.35in}

\catcode`@=12

\def\sect#1{\bigskip\medskip\goodbreak\noindent{\large\bf{#1}}\par\nobreak
        \medskip\markright{#1}}
\def\chsc#1#2{\phantom m\vskip.5in\noindent{\LARGE\bf{#1}}\par\vskip.75in
        \noindent{\large\bf{#2}}\par\medskip\markboth{#1}{#2}}
\def\Chsc#1#2#3#4{\phantom m\vskip.5in\noindent\halign{\LARGE\bf##&
        \LARGE\bf##\hfil\cr{#1}&{#2}\cr\noalign{\vskip8pt}&{#3}\cr}\par\vskip
        .75in\noindent{\large\bf{#4}}\par\medskip\markboth{{#1}{#2}{#3}}{#4}}
\def\chap#1{\phantom m\vskip.5in\noindent{\LARGE\bf{#1}}\par\vskip.75in
        \markboth{#1}{#1}}
\def\refs{\bigskip\medskip\goodbreak\noindent{\large\bf{REFERENCES}}\par
        \nobreak\bigskip\markboth{REFERENCES}{REFERENCES}
        \frenchspacing \parskip=0pt \renewcommand{\baselinestretch}{1}\small}
\def\unrefs{\normalsize \nonfrenchspacing \parskip=medskipamount}
\def\Item{\par\hang\textindent}
\def\Itemitem{\par\indent \hangindent2\parindent \textindent}
\def\makelabel#1{\hfil #1}
\def\topic{\par\noindent \hangafter1 \hangindent20pt}
\def\Topic{\par\noindent \hangafter1 \hangindent60pt}


\title{Two-Dimensional Dilaton Gravity in a Unitary Gauge}
{\AM}{92/16}{November 1992}
Reduced phase space formulation of CGHS model of 2d dilaton gravity
is studied in an
extrinsic time gauge. The corresponding Hamiltonian
can be promoted into a Hermitian operator acting in
the physical Hilbert space, implying a unitary evolution for the system.
Consequences for the black hole physics are discussed. In particular,
this manifestly unitary theory rules out the Hawking scenario for the
endpoint of the black hole evaporation process.
\endtitle

\sect{1. Introduction}

In a pioneering work \cite{cghs}, Callan, Giddings, Harvey and
Strominger have proposed a theory of two-dimensional dilaton
gravity coupled to matter
as a toy model for studying the formation, evaporation and back-reaction
of black
holes. The attractive features of the model are that it is classically exactly
solvable, it possesses black hole solutions and it is a renormalizible
field theory. The last feature raises a possibility that the corresponding
quantum theory may be tractable, and hence allow for the investigation of
the elusive issues associated with
the endpoint of the black hole evaporation \cite{hawk}. As shown by a series
of authors \cite{2dbh},
the solutions of the one-loop matter corrected equations of motion
are not free
of singularities, in contrast to the initial expectation by CGHS. Hawking
has even argued \cite{hawk2} that the solutions of any semi-classical
approximation scheme will be singular, suggesting that the possible
stabilization of the black hole by the quantum effects could be achieved only
if the gravitational field is quantised together with the matter fields.

Non-perturbative quantisation of the gravitational field in four spacetime
dimensions is still an
unsolved problem. However, in 2d, significant simplifications occur, most
notably the number of physical degrees of freedom of the gravitational field
is finite. In addition, the CGHS model is a renormalizible field theory.
However, the non-perturbative analysis is still a complicated problem.
Instead of using the path-integral techniques, one could try using the
canonical quantization methods, which were developed in the context of 4d
quantum gravity (for a review and references see \cite{asht}). In \cite{mik}
the canonical analysis of the model has been performed, an the Dirac type
quantisation investigated. It was shown that a set of non-canonical variables
can be found, forming an $SL(2,\bf{R})$ current algebra,
such that the constraints become quadratic in the new variables.
For a compact spatial manifold (i.e. circle) and picewise continious field
configurations, Fourier modes can be defined, and the physical Hilbert space
can be obtained from a cohomology of a Virasoro algebra. Although exactly
solvable, the configuration space of this model does not contain singular
solutions which can be associated with black holes. As suggested in \cite{mik},
a Schrodinger type equation would be more appropriate for quantizing a
more general
configuration space, which naturally leads one to employ the extrinsic time
variable approach \cite{kuch}.

In this paper we discuss the reduced phase space formulation of the
CGHS theory in an extrinsic time gauge. Our gauge fixing
conditions contain only the canonical variables, in contrast to the
usual gauge fixings, where the Lagrange multipliers are involved,
like the conformal gauge. Since we
are dealing with a reparametrization invariant system, a consistent
canonical gauge fixing must contain the definition of a time variable
\cite{mamik}. We construct a time variable $T(x,t)$, and in the gauge
$T(x,t) = t$ solve the constraints in terms of the independent canonical
variables. We obtain an explicit expression for the Hamiltonian of the
system in this gauge. That Hamiltonian can be promoted into a Hermitian
operator, acting on the physical Fock space, implying a unitary evolution.
Hence in this theory there are no anomalies associated with a non-unitary
evolution, like transitions from pure into mixed states, a pathology expected
at the endpoint of the black hole evaporation process \cite{hawk}.
However, it is still difficult to explicitly see what happens during
the gravitational collapse in this theory. This problem together with some
other caveats is discussed at the end of the paper.

\sect{2. Canonical Formulation}

The CGHS action \cite{cghs} is given by
$$ S= -\frac18 \int_{M} d^2 x \sqrt{-g}\left[ e^{-2\F}( R +
4 (\nabla \F)^2 + \l^2 ) + 4\su_{i=1}^N (\nabla\f_i)^2\right]
\quad,\eqno(2.1)$$
where $M$ is a 2d manifold,
$g\sb{\m\n}$ is a metric on $M$, $\F$ is a scalar field (dilaton),
$\l$ is a constant and $R$ is the 2d curvature scalar. $\f_i$ are
massless scalar fields, minimally coupled to gravity.
We will label the time coordinate $x^0 = t$ and the space coordinate
$x^1 = x$, while the corresponding derivatives will be denoted as $\dt{}$
and $'$, respectively.

Following the analysis in \cite{mik}, we perform the field
redefinitions \cite{tsey}
$$ \f = {1\over 4} e^{-2\F} \quad,\quad \tilde{g}_{\m\n}={4\f}
e^{-\f}g_{\m\n} \quad,\eqno(2.2)$$
so that the action becomes
$$ S= -\ha\int_{M} d^2 x \sqrt{-\tilde{g}}\left(
(\tilde{\nabla}\f)^2 + \tilde{R}\f + \frac14 \l^2 e^{\f}
+ \su_{i=1}^N (\tilde{\nabla}\f_i)^2 \right) \quad.\eqno(2.3)$$
The canonical formulation requires that the 2d manifold $M$ has a
topology of $\S \times {\bf R}$,
where $\S$ is the spatial manifold and ${\bf R}$ is the real line corresponding
to the time direction. $\S$ can be either a circle or a real line.
The compact spatial topology is relevant for cosmological solutions and
string theory, while the
non-compact spatial topology is relevant for 2d black holes.

After introducing the canonical momenta,
(2.3) takes the form \cite{mik}
$$ S= \int dt dx \left( p\dt{g} + \p\dt{\f} + \p^i\dt{\f_i} -
{{\cal N}\over\sqrt{g}}G_0 - n G\sb 1 \right) \quad, \eqno(2.4)$$
where we have omitted the tildas, $g =g_{11}$, $\cn$ and $n$ are the
laps and the shift vector and
$$\li{G\sb 0 (x) &=  - {2}g^2 p^2
-{2}gp\p + \ha(\f^{\prime})^2 + {\l^2\over 8}ge^{\f}
- {1\over2}{g^{\prime}\over g}\f^{\prime} + \f^{\prime\prime}+
\ha \su_{i=1}^N (\p_i^2 + (\f_i^{\prime})^2) \cr
G\sb 1 (x) &= \p\f^{\prime} - 2p^{\prime}g - pg^{\prime}+
\su_{i=1}^N \p_i\f^{\prime}_i \quad.&(2.5)\cr}$$

The constraints $G_0$ and $G_1$ form a closed Poisson bracket algebra
$$\li{ \{ G_0 (x), G_0 (y) \} &= -\d^{\prime} (x-y)(G_1 (x) + G_1 (y)) \cr
\{ G_1 (x), G_0 (y) \} &= -\d^{\prime} (x-y)(G_0 (x) + G_0 (y)) \cr
\{ G_1 (x), G_1 (y) \} &= -\d^{\prime} (x-y)(G_1 (x) + G_1 (y)) \quad,&(2.6)
\cr}$$
where the fundamental Poisson brackets are defined as
$$\{p(x),g(y)\}= \d (x -y)
\quad,\quad
\{\p(x),\f(y)\} = \d (x -y)\quad.\eqno(2.7)$$
$G_1$ generates the spatial diffeomorphisms, while $G_0$
generates the time translations of $\S$, in full analogy with the
4d gravity case.
Note that the algebra (2.6) is isomorphic to two comuting copies of the
one-dimensional
diffeomorphism algebra, which can be seen by redefining the constraints as
$$ T_{\pm} = \ha ( G_0 \pm G_1 ) \quad.\eqno(2.8)$$

Since we are dealing with a reparametrization invariant system, the
Hamiltonian vanishes on the constraint surface (i.e. it is proportional
to the constraints). Therefore the dynamics is determined by
the constraints only. Since $G_0$ and $G_1$ are independent, there will be
$(2+N) - 2 = N$ local physical degrees of freedom. When $N=0$, there is
only a finite number of global physical degrees of freedom (zero modes
of $g$ and $\f$), and one is dealing with
some kind of a topological field theory. When $N\ne 0$, these global
degrees of freedom
will be present, together with the local ones.

The variables $(g,p,\f,\p)$ are not
convenient for quantization, since $G_0$ is a non-polynomial function of
these variables. First we perform a canonical transformation in order to
get rid off the $e^{\f}$ term in $G_0$
$$ \li{ g = e^{-\tilde\f}\tilde g \quad,&\quad p = e^{\tilde\f} \tilde p
\cr \f = \tilde \f \quad,&\quad \p = {\tilde \p} +  \tilde p \tilde g
\quad.&(2.9)\cr}$$
The constraints now become
$$\li{G\sb 0 (x) &=  - 4g^2 p^2
-2gp\p + (\f^{\prime})^2 + \L g
- \ha {g^{\prime}\over g}\f^{\prime} + \f^{\prime\prime} +
\ha\su_{i=1}^N (\p_i^2 + (\f_i^{\prime})^2)\cr
G\sb 1 (x) &= \p\f^{\prime} - 2p^{\prime}g - pg^{\prime} + \su_{i=1}^N
\p^i\f_i^{\prime} \quad,&(2.10)\cr}$$
where we have dropped the tildas and $\L = {\l^2\over8}$. Now it is
convenient to define
the $SL(2,{\bf R})$ variables introduced in \cite{mik}
$$\li{J\sp + &= -{\sqrt{2}\over 2{g}} T\sb - + {\L\over2\sqrt2}\cr
J\sp 0 &= gp + {1\over4}\left( \p -
{1\over2}{g^{\prime}\over g}\right)  \cr
J\sp - &= {1\over\sqrt{2}} g \quad,&(2.11)\cr}$$
and a $U(1)$ current
$$ P_D = {1\over2}\left( \p -
{1\over 2} {g^{\prime}\over g} + 2\f^{\prime} \right)
\quad.$$
The $(J^a ,P_D)$ variables satisfy an
$SL(2,{\bf R})\otimes U(1)$ current algebra
$$\li{ \{ J^a (x), J^b (y)\} &= f\sp{ab}\sb c J^c (x)
\d (x-y) - {1\over4} \h^{ab}\d^{\prime}(x-y) \cr
\{ P_D (x) , P_D (y) \} &= - \d^{\prime}(x-y)\quad,&(2.12)\cr}$$
where $f\sp{ab}\sb c = 2\e\sp{abd}\h\sb{dc}$ with
$\h^{+-}=\h^{-+}=2$, $\h^{00}=-1$,
and $\{J, P_D\} =0$. Instead of using the canonical variables $(\p_i,\f_i)$,
we introduce the left/right moving currents
$$P_i =  {1\over\sqrt{2}}\left( \p_i + \f_i^{\prime}\right) \quad,\quad
\tilde{P}_i =  {1\over\sqrt{2}}\left( \p_i - \f_i^{\prime}\right) \quad,
\eqno(2.13)$$
satisfying
$$\{ P_i (x) , P_j (y) \} = -\d_{ij} \d^{\prime}(x-y)\quad,\quad
\{ \tilde{P}_i (x) , \tilde{P}_j (y) \} = \d_{ij} \d^{\prime}(x-y)\quad,
\eqno(2.14)$$
and $\{ P, \tilde{P} \} = 0$.
Now one can show that the energy-momentum tensor
associated to the algebra (2.12) via the Sugavara construction
$$ {\cal S} = 2\h_{ab}J^a J^b - (J^0)^{\prime}+ \ha P_D^2 + \ha P_D^{\prime}
+ \ha \su_{i=1}^N P_i^2 \quad, \eqno(2.17)$$
satisfies $ {\cal S} \approx T_+ $.
Therefore the constraints become
$$ J^+ (x) -\m = 0 \quad,\quad
{\cal S}(x) = 0\quad, \eqno(2.18)$$
where $\m = {\L\over 2\sqrt2}$.

Now it is convenient to
introduce three new variables $\b(x)$, $\g(x)$ and $P\sb L (x)$ \cite{mik}
such that
$$ \li{J^+ &= \b\cr
J^0  &=-\b\g - {1\over 2} P\sb L \cr
J^-  &= \b\g^2 +  \g P\sb L  - \ha \g^{\prime} \quad,&(2.18)\cr}$$
where
$$ \{\b (x) ,\g (y) \} = - \d (x-y) \quad,\quad
\{ P\sb L (x) , P\sb L (y)\} = \d^{\prime} (x-y)
\quad,\eqno(2.19)$$
with the other Poissons brackets being zero. Then the expressions (2.18)
satisfy the $SL(2,{\bf R})$ current algebra (2.12), and represent the
classical analogue of the Wakimoto transformation \cite{wak}.
The ${\cal S}$ constraint
then becomes
$$ {\cal S} = \b^{\prime}\g - \ha P^2_L  +
\ha P^{\prime}_L + \ha P^2_D + \ha P^{\prime}_D
+ \ha \su_{i=1}^N P_i^2  = 0 \quad.\eqno(2.20)$$
If we define $B(x) = \b(x) - \l$ and $\G (x) = \g(x)$, then the $J^+$
constraint implies that $B =0$, and consequently we can omitt the canonical
pair $(B,\G)$ from the theory. Therefore we are left with $P_L , P_D$ and
$P_i$ variables, obeying only one constraint
$$ 2{\cal S} = - P_L^2 +  P_L^{\prime}
 + P^2_D + P^{\prime}_D + \su_{i=1}^N P_i^2 =0
\quad.\eqno(2.21)$$

The form of the Poisson brackets of $P_L$ and $P_D$ allow us to introduce
a canonical pair $(P(x) , T(x))$ such that
$$ P_L = {1\over\sqrt2}(P - T^{\prime} ) \quad,\quad P_D = {1\over\sqrt2}(P +
T^{\prime} ) \quad.\eqno(2.22)$$
Note that the definition (2.22) implies that the zero-mode parts of
$P_L$ and $P_D$ are equal. When $N =0$, this is true on the
constraint surface, but away from the constraint surface these zero
modes are independent. Therefore we are going to modify the eq. (2.22) by
introducing an independent zero-mode momentum $p$ such that
$$ P_L = {1\over\sqrt2}(P - T^{\prime} ) \quad,
\quad P_D = p + {1\over\sqrt2}(P + T^{\prime} ) \quad.\eqno(2.23)$$
Then the ${\cal S}$ constraint becomes
$$ {\cal S}= (p + \sqrt2 T^{\prime})(p + \sqrt2 P) + \sqrt2 P^{\prime}
+ \su_{i=1}^N P_i^2 =0 \quad.
\eqno(2.24)$$
Now one can easily solve the eq. (2.24) for $T$ or $P$,
and therefore put $\cs$ into form which is linear in one of the
momenta, a step which is crucial for formulating a Schrodinger type equation
\cite{kuch}. Although in this way one preserves the manifest diffeomorphism
covariance, the corresponding multifingered time Schrodinger equation is
difficult to solve. Instead, we fix the time reparametrization invariance
by choosing the gauge
$$ T(x,t)= t \quad.\eqno(2.25)$$
Then from the eq. (2.24) we get
$$ P(x) = -{p\over\sqrt2} - {1\over\sqrt2}e^{-px} \int^x dy \,
e^{py}\,  \su_{i=1}^N P_i^2 (y)
\quad.\eqno(2.26)$$
Hence the independent canonical variables are $(\p_i(x) ,\f_i(x))$
together with the $x$-independent variables
$(p,q)$. The $(p,q)$ variables are the global remnants of the
graviton-dilaton sector, and they represent the physical degrees of
freedom of that sector.
The Hamiltonian for the independent canonical
variables can be deduced from the $\int dx P\dt{T}$ part of the action to be
$$ H = {cp\over\sqrt2} + {1\over\sqrt2}\int_{-\infty}^{\infty} dx
e^{-px} \int^x dy\, e^{py}\, \su_{i=1}^N P_i^2 (y) \quad,\eqno(2.27)$$
where $c$ is a constant. In the compact case $c$ is proportional to
the volume of $\S$, and can be set to $1$. In the non-compact case,
the value of $c$ can be determined from the requirement of the
asymptotic flatness of
the black-hole solution, whose ADM mass is asymptotically conserved
energy \cite{wit}, and therefore $M=H=cp$.

The formulas (2.26) and (2.27) simplify if we use the Fourier modes of
$P_i$
$$ P_i (x) = {1\over\sqrt{2\p}}\int_{-\infty}^{\infty} dk\, e^{ikx}\,
\a^i_k  \quad,\eqno(2.28)$$
and analogously for $\tilde{P}_i$. In particular one gets for the
Hamiltonian (2.27)
$$ H = {cp\over\sqrt2} + {1\over 2\sqrt{2}p} \int_{-\infty}^{\infty}
dk\, \a_{-k}^i \a_k^i
\quad, \eqno(2.29)$$
which is almost like a free-field Hamiltonian, except for the
non-polynomial dependence on the momentum $p$.

Note that $H$ would be positive definite if $p$ was restricted to ${\bf R}_+$.
It is a non-trivial task to deduce directly from our approach what is
the range of $p$, but
when compared to the results of the Dirac analysis \cite{mik}, $p$ can be
identified with the energy of a free relativistic 2d particle, whose
range is positive.
There is also a disconnected piece, corresponding to the negative energies.
One obtains a similar result for the reduced configuration space of the gravity
plus dilaton sector in the $V(\f)=const.$ model,
i.e. two disconnected ${\bf R}_+$ spaces \cite{nsa}. Both models have the
same reduced phase space since the constraints can be brought into
an identical form \cite{mik}.

\sect{3. Quantization}

Given the reduced phase space and the corresponding Hamiltonian, one can
define the quantum theory by the Schrodinger equation
$$ i\pa_t\Psi = \hat{H}\Psi \quad,\eqno(3.1)$$
where $\hat{H}$ is an operator corresponding to the classical expression
(2.29). $\Psi$ belongs to a Hilbert space constructed from the canonical
algebra of the basic variables $(p,q, \p_i(x), \f_i(x) )$, which are now
promoted into hermitian operators. As in the classical case, it is convenient
to use the $P_i$ and $\tilde{P}_i$ operators, satisfying
$$ [P_i (x),P_j (y)] = -i\d^{\prime}(x-y)\d_{ij}\quad,\quad
[\tilde{P}_i (x),\tilde{P}_j (y)] = i\d^{\prime}(x-y)\d_{ij}
\quad,\eqno(3.2)$$
and $[P_i,\tilde{P}_j] = 0$, while for $p$ and $q$ we will take
$$[p,q]=ip \quad,\eqno(3.3)$$
since $p\in {\bf R}_+$. Given the relations (3.2) there is
an immediate problem of how to order the $P$'s in the expression (2.29).
However, given the simple form of $H$ in terms of the $\a$
modes, and the fact that they resemble particle creation and anhilation
operators, we can define a quantum theory based on the Hilbert space
$$ \ch^* = \ch (p) \otimes \cf (\a ) \otimes \cf (\tilde{\a}) \eqno(3.4)$$
where $\ch (p)$ is the Hilbert space associated with the $(p,q)$
algebra, while $\cf (\a )$ and $\cf (\tilde{\a})$ are the Fock spaces built
on the vacuum
$$ \a^i_k \sket{0}= \tilde{\a}^i_{-k} \sket{0}= 0 \quad,\quad k \ge 0
\quad.\eqno(3.5)$$
One can now introduce the standard field-theory creation and anhilation
operators as
$$\li{ a_i(k) =& {1\over\sqrt{k}}\a^i_k \quad,\quad k >0 \cr
       a_i(k) =& {1\over\sqrt{|k|}}\tilde{\a}^i_{k} \quad,\quad k <0
\quad. &(3.6) \cr}$$
Therefore $\a_k$ corresponds to the right-moving ($k>0$) quant, while
$\tilde{\a}_k$ corresponds to the left-moving ($k<0$) quant.

Given the Hilbert space $\ch^*$, the hamiltonian $H$ can be promoted into a
hermitian  operator
$$ \hat{H} = {cp\over\sqrt2} + {1\over\sqrt2 p} \int_{0}^{\infty}
dk|k| a_i^{\dagger}(k)a_i (k) \quad. \eqno(3.7)$$
Note the absence of the left-moving modes in the expression (3.7).
This is the consequence of the fact that the $\cs$ constraint does not
depend on the $\tilde{P}_i$ variables. This asymmetry arises from our
choice of the variables and the gauge-fixing procedure. In (2.11) we set $J^+
\approx T_-$ and subsequently $\cs \approx T_+$. Then we solve the $J^+$
constraint by setting $\b = \m$ while the $\cs$ constraint is solved by
chosing the gauge (2.25), which is a choice of the time variable and
therefore the $\cs$ constraint is transformed into a Schrodinger equation.
Hence in the gauge (2.25) the $T_+$ constraint generates the
time translations, while $T_-$ generates the spatial diffeomorphisms
and consequently $\tilde{P}_i$ are frozen
(integrals of motion). Clearly our choice of the variables and the gauge is
convenient for describing a one-sided collapse, i.e. when initially one
has only a right-moving matter.

\sect{4. Concluding Remarks}

The Hamiltonian of our theory is a Hermitian operator in the physical
Hilbert space, and
therefore any time evolution will be unitary. This implies in
particular that there will not be any transitions from pure into mixed
states, which rules out the Hawking scenario \cite{hawk} for the
endpoint of the black hole evaporation process. However, in order to
to see what really happens during the gravitational collapse one has
to carefully study the black-hole solutions in this theory. The
spatial metric $g(x)$ can be written as
$$ g(x) = {\l^2\over 16} \g^2 (x) - {\g(x)\over\sqrt2}\left( p + \frac1{2\p}
\int_{-\infty}^{+\infty} dk \, L_k\, {e^{ikx}\over p + ik} \right) -
\sqrt2\g^{\prime}(x) \quad,\eqno(4.1)$$
where
$$ L_k = \int_{-\infty}^{+\infty} dq\, \a^i_{k-q}\a^i_q \quad.\eqno(4.2)$$
Classical equations of motion imply
$$ \dt{p}= \{ H,p \} = 0 \quad,\quad \dt{\a}_k = \{ H, \a_k\} ={k\over
2\sqrt2 p}\a_k
\quad,\eqno(4.3) $$
so that from the eq. (4.1) one can find the spatial metric at any
time. Note the
similar structure of the expression (4.1) and the corresponding
expression of CGHS \cite{cghs}, where our $p$ is analogous to their
$M$ (ADM mass of the black hole), our arbitrary gauge function $\g(x)$
is analogous to their $w_{\pm}(x)$ gauge functions,
and the dependence on the scalar fields is similar. The differences
come from the fact that we are working in some type of the Polyakov
light-cone gauge \cite{poly}, while CGHS are in the conformal gauge.

When $N=0$, then the black-hole solution is equivalent to a choice of
$\g(x)$ such that
$$ {\l^2\over 16} \g^2 (x) - {\g(x)\over\sqrt2} p
-\sqrt2\g^{\prime}(x) = {e^{\l x}\over 1 - {M\over\l} e^{-\l x}}
\quad.\eqno(4.4)$$
Solutions of this equation exist; however, we could not find an
explicit expression. Such an expression will give the relation
between the parameters $p$ and $M$, and it will constitute an
independent check of $M=cp$.

Since $H\approx p$ in the $N=0$ case, one has an eternal
black hole. Clearly in order to get some interesting effects, $N$ must
be different from zero. Then $g(x)$ is given by the expression (4.1),
which becomes an operatorial expression in the quantum theory.
A normal ordering ambiguity in the $L_k$ operators then appears. The standard
normal ordering prescription causes a c-number anomaly in the
comutator $[\hat{g}(x),\hat{g}(y)]$, and we believe that this is a
technical problem which could be overcomed by appropriate
modifications. Note that in the Dirac approach a $c$-number anomaly
appears in the diffeomorphism algebra \cite{mik}. It would be
interesting to see whether this anomaly is in any way equivalent to
the metric anomaly in our approach.
A more difficult problem is the construction of a
hermitian operator associated with the scalar curvature $R$. This operator
is important since it will give a measure of a
singularity. $R$ is certain to be a non-polynomial function of the $p$
and the $a(k)$'s, which will be the main source of difficulties in
constructing the $\hat{R}$ operator.

An important issue which has to be analyzed is the Hawking effect. A
natural way to do this in our model is to construct a state $\sket{\j_0}$
such that
$$\hat{g}(x)\sket{\j_0} = g_{reg}(x)\sket{\j_0}\quad,\eqno(4.5)$$
where $g_{reg}(x)$ is a non-singular metric. Then evolve $\sket{\j_0}$ in time
by the evolution operator $e^{-i\hat{H}t}$. At some time $t_A$ an
apparent horizon will form in the effective metric
$\sbra{\j_0}e^{i\hat{H}t}\hat{g}e^{-i\hat{H}t}\sket{\j_0}$.
Then a reduced density matrix $\hat{\r}$ can be introduced, by tracing
out the states
which are beyond the horizon \cite{{hawk},{wald}}. How to define
these states is not clear at the moment, but when this problem is resolved
then one could in principle answer the
questions about the thermal nature of $\hat{\r}$, i.e. when
$$\hat{\r} \approx {1\over Z}e^{-\b \hat{H}}\quad,\eqno(4.6)$$
and what are the non-perturbative corrections to the Hawking temperature
$$\b = {4\p \over \l} + .... \quad.\eqno(4.7)$$
Moreover, by analyzing the effective scalar curvature
$$R_{eff}(x,t)= \sbra{\j_0}e^{i\hat{H}t}\hat{R}(x)e^{-i\hat{H}t}\sket{\j_0}$$
one should be able to say what happens with the singularity. Ideally,
$R_{eff}(x,t)$ should stay a regular function for any $t$.

We should emphasize that in our quantization scheme the topology of
the space-time stays fixed. One could argue that this is the main
reason why the theory is unitary, and no violations of quantum
mechanics occur. This may well be the case, and we should point out
that in the context of 2d gravity introduction of the topology change is
equivalent to introducing the interactions in the corresponding string
field theory. As a result, the string field $\Psi$ will not satisfy the
linear equation (3.1), but instead the eq. (3.1) will be modified by $\Psi^2$
and higher order terms. This will directly violate the quantum
mechanics. Formally, one can invoke the third quantization in order to
get around this problem; however, it is not clear how to define such
a theory. Matrix models approach offers a definition \cite{matrix}, and
there are indications that the Wheeler-DeWitt equation is not
satisfied \cite{mwdw}.

Clearly a lot of work remains to be done in order to answer all these
questions. The main difficulty at the moment are the explicit
calculations. However,
we belive that the reduced phase-space quantization
approach can answer, at least qualitatively, some of the issues raised
in our discussion. Study of
the theory in other gauges should also be beneficial, since then the
questions of the diffeomorphism invariance could be analyzed and
the results in different gauges compared.


\begin{thebibliography}{99}

\bibitem{cghs} C.G. Callan, S.B. Giddings, J.A. Harvey and A. Strominger,
\PRD 45 (1992) R1005

\bibitem{hawk} S.W. Hawking, \PRD 14 (1976) 2460

\bibitem{2dbh} T. Banks, A. Dabholkar, M.R. Douglas and M. O'Loughlin,
\PRD 45 (1992) 3607;\\
J.G. Russo, L. Susskind and L. Thorlacius, \PL 292B (1992) 13

\bibitem{hawk2} S.W. Hawking, {\it Evaporation of two-dimensional black holes},
Caltech preprint CALT-68-1774 (1992)

\bibitem{tsey} J.G. Russo and A.A. Tseytlin, \NP 382 (1992) 259

\bibitem{asht} A. Ashtekar, Nonperturbative Canonical Quantum Gravity,
World Scientific, Singapore (1991);\\
C. Isham, {\it Conceptual and Geometrical Problems in Quantum Gravity},
Imperial preprint, Imperial/TP/90-91/14 (1991)

\bibitem{mik} A. Mikovi\'c, {\it Exactly Solvable Models of 2d Dilaton Quantum
Gravity}, Queen Mary preprint, QMW/PH/92/12 (1992)

\bibitem{wak} M. Wakimoto, Commun. Math. Phys. 104 (1986) 605

\bibitem{nsa} J. Navarro-Salas, M. Navarro and V. Aldaya, {\it Covariant phase-
space quantization of induced 2d gravity}, Cern preprint, CERN-TH.6537/92
(1992)

\bibitem{kuch} K. Kuchar, Quantum Gravity 2: A second Oxford Symposium, eds.
C.J. Isham, R. Penrose and D.W. Sciama, Clarendon Press, Oxford (1981)

\bibitem{wald} R.M. Wald, Commun. Math. Phys. 45 (1975) 9

\bibitem{wit} E. Witten, \PRD 44 (1991) 314

\bibitem{poly} A. Polyakov, Mod. Phys. Lett. A2 (1987) 893

\bibitem{mamik} N. Manojlovi\'c and A. Mikovi\'c, \NP 382 (1992) 148

\bibitem{matrix} E. Brezin, C. Itzykson, G. Parisi and J.B. Zuber,
Commun. Math. Phys. 59 (1978) 35;\\
M.R. Douglas and S.H. Shenker, \NP 335 (1990) 635;\\
E. Brezin and V. Kazakov, \PL 236B (1990) 144;\\
D. Gross and A. Migdal, Phys. Rev. Lett. 64 (1990) 127

\bibitem{mwdw} A. Cooper, L. Susskind and L. Thorlacius, {\it The
classical limit of quantum gravity isn't}, SLAC preprint,
SLAC-PUB-5413 (1991)

\end{thebibliography}
\end{document}